\documentclass[twoside,11pt]{article}

\usepackage{jmlr2e}
\usepackage{listings}
\usepackage[utf8]{inputenc}

\usepackage{listings}
\usepackage{xcolor}
\usepackage{amsmath}

\definecolor{codegreen}{rgb}{0,0.6,0}
\definecolor{codegray}{rgb}{0.5,0.5,0.5}
\definecolor{codepurple}{rgb}{0.58,0,0.82}
\definecolor{backcolour}{rgb}{0.95,0.95,0.92}

\lstdefinestyle{mystyle}{
    backgroundcolor=\color{backcolour},   
    commentstyle=\color{codegreen},
    keywordstyle=\color{magenta},
    numberstyle=\tiny\color{codegray},
    stringstyle=\color{codepurple},
    basicstyle=\ttfamily\footnotesize,
    breakatwhitespace=false,         
    breaklines=true,                 
    captionpos=b,                    
    keepspaces=true,                 
    numbers=left,                    
    numbersep=5pt,                  
    showspaces=false,                
    showstringspaces=false,
    showtabs=false,                  
    tabsize=2
}

\ShortHeadings{TensorDiffEq}{McClenny, Haile, and Braga-Neto}
\firstpageno{1}

\begin{document}

\title{\textit{TensorDiffEq}: Scalable Multi-GPU Forward and Inverse Solvers for Physics Informed Neural Networks}

\author{\name Levi D. McClenny\email levimcclenny@tamu.edu \\
       \addr Department of Electrical Engineering\\
       Texas A\&M University\\
       College Station, TX 77840, USA
       \AND
       \name Mulugeta A. Haile \email mulugeta.a.haile.civ@mail.mil \\
       \addr US Army CCDC Army Research Lab\\
       Aberdeen Proving Ground \\
       Aberdeen, MD, USA
       \AND
       \name Ulisses M. Braga-Neto \email  ulisses@tamu.edu\\
       \addr Department of Electrical Engineering\\
       Texas A\&M University\\
       College Station, TX 77840, USA}

\maketitle

\begin{abstract}%   <- trailing '%' for backward compatibility of .sty file
Physics-Informed Neural Networks promise to revolutionize science and engineering practice, by introducing domain-aware deep machine learning models into scientific computation. Several software suites have emerged to make the implementation and usage of these architectures available to the research and industry communities. Here we introduce\linebreak TensorDiffEq, built on Tensorflow 2.x, which presents an intuitive Keras-like interface for problem domain definition, model definition, and solution of forward and inverse problems using physics-aware deep learning methods. TensorDiffEq takes full advantage of Tensorflow 2.x infrastructure for deployment on multiple GPUs, allowing the implementation of large high-dimensional and complex models. Simultaneously, TensorDiffEq supports the Keras API for custom neural network architecture definitions. In the case of smaller or simpler models, the package allows for rapid deployment on smaller-scale CPU platforms with negligible changes to the implementation scripts. We demonstrate the basic usage and capabilities of TensorDiffEq in solving forward, inverse, and data assimilation problems of varying sizes and levels of complexity. The source code is available at \url{https://github.com/tensordiffeq}.\footnote{Package documentation is available at \url{https://docs.tensordiffeq.io}}
\end{abstract}

\begin{keywords}
  Scientific Machine Learning, PINNs, Neural PDEs, Scientific Computation, Numerical Methods, Neural Networks, Physics-Informed Deep Learning
\end{keywords}

\section{Introduction}

As part of the burgeoning field of scientific machine learning~\citep{osti_1478744}, physics-informed neural networks (PINNs) have emerged recently as an alternative to traditional partial different equation (PDE) solvers~\citep{raissi2019physics,raissi2018forward,wight2020solving,wang2020and}, and have given rise to the larger field of study in neural network approximation of PDE systems, generally referred to as Neural PDEs. Typical black-box deep learning methodologies do not take into account the underlying physics of the problem domain. 
The Neural PDE approach is based on constraining the output of a deep neural network to satisfy a physical model specified by a PDE. PINNs typically perform this task via PDE-constrained regularization of a residual function defined by the approximation of the solution network \verb|u| and forward-pass calculations through the physics of the PDE model, with the applicable derivatives of \verb|u| calculated via reverse-mode automatic differentiation in a modern deep learning framework such as Tensorflow~\citep{abadi2016tensorflow}.

The potential of using neural networks as universal function approximators to solve PDEs had been recognized since the 1990's~\citep{dissanayake1994neural, lagaris1998artificial}. However, Physics-Informed Neural Networks promise to take this approach to a different level through deep neural networks, the exploration of which is now possible due to the vast advances in computational capabilities and training algorithms since that time~\citep{abadi2016tensorflow, revels2016forward} and modern congenial automatic differentiation software~\citep{baydin2017automatic,paszke2017automatic}. 

A great advantage of the PINN architecture over traditional time-stepping PDE solvers is that the entire spatial-temporal domain can be solved at once using collocation points distributed quasi-randomly (rather than on a grid) across the spatial-temporal domain, in a process that can be massively parallelized via GPU. As we have continued to see GPU capabilities increase in recent years, a method that relies on parallelism in training iterations could begin to emerge as the predominant approach in scientific computing. To this end, while other software suites exist to define and solve PINNs~\citep{DifferentialEquations.jl-2017, lu2021deepxde, hennigh2020nvidia, haghighat2021sciann}, many of those platforms are either restricted to single-GPU implementation or are not fully open-source. 
Additionally, with full support and customization capabilities of the Keras neural network ecosystem built in to the package, researchers and practitioners can define and train their own custom neural network architectures to approximate the solution of their problem domains.
 \textit{TensorDiffEq} provides these scalable, modular, and customizable multi-GPU architectures and solvers in a fully open-source platform, tapping into the collective intelligence of the field to improve the implementation of the software and provide input on the direction, structure, and feature coverage of the framework.

\section{Mathematical Underpinnings of PINNs}

Consider a general nonlinear PDE of the form:
\begin{align}
    & \mathcal{N}_{\boldsymbol{x},t}[ u(\boldsymbol{x}, t)] = 0\,, \ \ \boldsymbol{x} \in \Omega\,, \ t \in [0,T]\,, \\
    & u(\boldsymbol{x}, t) = g(\boldsymbol{x},t)\,, \ \  \boldsymbol{x} \in \partial\Omega, \ t \in [0,T]\,,\label{eq:boundary} \\
    & u(\boldsymbol{x}, 0) = h(\boldsymbol{x})\,, \ \
    \boldsymbol{x} \in \Omega\,, \label{eq:initial}
\end{align}
where $\boldsymbol{x} \in \Omega$ is a spatial vector variable in a domain $\Omega \subset R^d$, $t$ is time, and $\mathcal{N}_{\boldsymbol{x},t}$ is a spatial-temporal differential operator. Following~\cite{raissi2019physics}, let $u(\boldsymbol{x}, t)$ be approximated by the output $u(\boldsymbol{x},t;\boldsymbol{w})$ of a deep neural network with inputs $\boldsymbol{x}$ and $t$. Define the residual network $r(\boldsymbol{x},t;\boldsymbol{w})$, which share the same network weights $\boldsymbol{w}$ as the approximation network $u(\boldsymbol{x},t;\boldsymbol{w})$, and satisfies: 
\def\bx{\boldsymbol{x}}
\def\bw{\boldsymbol{w}}
\begin{equation}
r(\boldsymbol{x},t;\bw) := \mathcal{N}_{\bx,t}[u(\boldsymbol{x},t;\bw)]\,, \label{eq:res_loss}
\end{equation}
where all partial derivatives can be computed by automatic differentiation 
methods~\citep{baydin2017automatic,paszke2017automatic}. The shared network weights $\boldsymbol{w}$ are trained by minimizing a loss function that penalizes the output for not satisfying (1)-(3): \begin{equation}
    \mathcal{L}(\boldsymbol{w}) = \mathcal{L}_s(\boldsymbol{w}) + \mathcal{L}_r(\boldsymbol{w}) + \mathcal{L}_{b}(\boldsymbol{w}) + \mathcal{L}_{0}(\boldsymbol{w})\,, \label{eq:loss}
\end{equation}
where $\mathcal{L}_s$ is the loss corresponding to sample data (if any), $\mathcal{L}_r$ is the loss corresponding to the residual (\ref{eq:res_loss}), $\mathcal{L}_{b}$ is the loss due to the boundary conditions (\ref{eq:boundary}), and $\mathcal{L}_{0}$ is the loss due to the initial conditions~(\ref{eq:initial}):
\begin{align}
    \mathcal{L}_s(\bw) &= \frac{1}{N_s}\sum^{N_s}_{i=1} |u(\boldsymbol{x}^i_s, t^i_s;\bw) - y^i_s|^2, \\
    \mathcal{L}_r(\bw)&= \frac{1}{N_r}\sum^{N_r}_{i=1} r(\boldsymbol{x}^i_r, t^i_r;\bw)^2, \\
    \mathcal{L}_{b}(\bw) &= \frac{1}{N_b}\sum^{N_b}_{i=1}|u(\boldsymbol{x}^i_b, t^i_b;\bw) - g^i_b|^2, \, \label{Lb}\\
    \mathcal{L}_{0}(\bw) &= \frac{1}{N_0}\sum^{N_0}_{i=1}|u(\boldsymbol{x}^i_0,0;\bw)- h_0^i|^2.
\end{align}
where $\{\boldsymbol{x}_s^i, t_s^i, y_s^i\}_{i = 1}^{N_s}$ are sample data (if any), $\{\boldsymbol{x}_0^i,  h_0^i=h(\boldsymbol{x}_0^i)\}_{i = 1}^{N_0} $ are initial condition point,  $\{\boldsymbol{x}_b^i, t^i_b, g_b^i = g(\boldsymbol{x}_b^i, t^i_b))\}_{i = 1}^{N_b}$ are boundary condition points, $\{\boldsymbol{x}_r^i, t^i_r\}_{i = 1}^{N_r} $ are collocation points randomly distributed in the domain $\Omega$, and $N_s, N_0, N_b$ and $N_r$ denote the total number of sample data, initial points, boundary points, and collocation points, respectively. 
%In a mini-batch setting, these can be modified to be the size of the batch %$N_{batch}$. 
The network weights $\bw$ can be tuned by minimizing the total training loss $\mathcal{L}(\bw)$ via standard gradient descent procedures used in deep learning.  

\section{Using \textit{TensorDiffEq} for Forward Problems}

\textit{TensorDiffEq} has a boilerplate model that can loosely be followed in most instances of usage of the package. In forward problems, this process is generally described in the following order:

\begin{enumerate}
    \item Define the problem domain
    \item Describe the physics of the model 
    \item Define the Initial Conditions and Boundary Conditions (IC/BCs)
    \item Define the neural network architecture
    \item Select and define the solver
    \item Solve the PDE using the \verb|fit| method
\end{enumerate}

Each of these steps has multiple options and definitions in the \textit{TensorDiffEq} solution suite. The following sections will provide a brief overview of some of the built-in functionality of the package. 

\subsection{Define the Problem Domain}

A \verb|Domain| object is the first essential component of defining a problem in \textit{TensorDiffEq}. The domain object contains primitives for defining the problem scope used later in your definitions of boundary conditions, initial conditions, and eventually to sample collocation points that are fed into the PINN solver.

The \verb|Domain| object is defined iteratively. As many dimensions as are required can simply be added to the domain using the \verb|add| method. This means \textit{TensorDiffEq} can be used to solve spatial (steady-state) or spatiotemporal 2D, 3D, or $N$D problems. 

% The \verb|Domain| object also contains support to sample the ND domain described in the problem via Latin-Hypercube Sampling, and no input from the user is required to sample the domain. The LHS-sampled points live in the \verb|Domain| object and are automatically used in solving the PDE via a collocation method, if the collocation method solver is selected. 

\subsection{Describe the Physics of the Model}

Since \textit{TensorDiffEq} is built on top of Tensorflow~\citep{tensorflow2015-whitepaper} physics of the model can be defined via a strong-form PDE, with gradients defined using the built-in \verb|tf.gradients| function. This allows for a definition similar to that seen in \cite{raissi2019physics}. An example of defining the PDE for a viscous Burger's system is shown below:

\begin{lstlisting}[language=Python]
def f_model(u_model, x, t):
    u = u_model(tf.concat([x, t], 1))
    u_x = tf.gradients(u, x)
    u_xx = tf.gradients(u_x, x)
    u_t = tf.gradients(u, t)
    f_u = u_t + u * u_x - (0.01 / tf.constant(math.pi)) * u_xx
    return f_u
\end{lstlisting}

Due to the nature of how the PDE system is defined in \textit{TensorDiffEq}, one could define a separate system of \verb|u| and allow for a coupled PDE definition using a similar style as the one shown above. 

\subsection{Define the ICs/BCs}
\textit{TensorDiffEq} supports various types of ICs and BCs and the list will continue to grow. The ICs and BCs that require functions allow for intuitive definitions of those functions of system variables as a Python \verb|function|, which allows for nonlinear and non-continuous function definitions of state variables. One could define piece-wise functions, Boolean functions, etc using this verbiage and it would be valid input to \textit{TensorDiffEq}'s solvers. At the time of this writing, \textit{TensorDiffEq} supports constant Dirichlet, Function Dirichlet, and periodic BCs, as well as function-based ICs. 
\textit{TensorDiffEq} takes the ICs and BCs as a list, therefore one can add as many as necessary to define the system. 
%Furthermore, it allows for some level of Deep Learning-based approximation in the event a BC is not known.
If a BC is not defined on a particular boundary or it is overlooked in the problem definition then the solver will attempt to approximate that boundary using PDE-constrained regularization of the inner points on or around that boundary.

\subsection{Define the Neural Network Architecture}
The default architecture of the neural network is a fully connected MLP defined in the Keras API~\citep{chollet2015keras}. To take advantage of the built-in MLP, a list of hidden layer sizes is passed into the solver. However, this baseline architecture can be overwritten by any Keras neural network. Currently, the solver requires the number of inputs of the neural network to be the same as the number of dimensions of the system, and the output is the scalar value of the approximation of $u(\textbf{X})$ at that combination of input points. However, this ``single-network'' output architecture is actively being expanded at the time of this writing. 

In the event one desires to add batch norm, residual blocks, etc, then the Keras API could be used to define the model and the internal parameters of \textit{TensorDiffEq's} solvers could be modified to use that network as the solution network for $u(\textbf{X})$. In this way, so long as the input to the neural network has the correct dimensionality for the system (i.e. 3 nodes for a problem with \verb|x, y, t| dimensions) and the output node is the correct number of dimensions then one could build any architecture the Keras API allows and pass it into the solver. This features also allows for custom neural network layer support using the Keras \verb|lambda| layer ecosystem, allowing for complete autonomy in the definition of the neural network model internals and training via built-in Keras optimizers. 

\subsection{Select and Define The Solver}
\textit{TensorDiffEq} is a suite designed to provide forward and inverse PINN solvers. As such, there are various solvers to perform these tasks. At the time of this writing, there are \textit{Collocation Method} solvers for forward modeling and the \textit{Discovery Model} for inverse modeling. 

Hyperparameter selection can be modified by the user by overwriting the default Adam optimizer~\citep{kingma2014adam} with any of the other available optimizers in Keras, to include AdaDelta~\citep{zeiler2012adadelta}, Root-Mean-Square Propagation, SGD, and others. Some of these different optimization techniques prove more stable in training than others, and there exist various methods of modifying the loss function of the collocation solver to improve convergence~\citep{wang2020and, wang2020understanding}. To this end, \textit{TensorDiffEq} supports self-adaptive training methods, which have proven to be effective in helping semi-linear PDE systems, such as Allen-Cahn~\citep{allen1972ground}, converge where the baseline collocation method fails~\citep{mcclenny2020self}. Other methods of improving convergence in Neural PDE and PINN training are continuously being considered. 

\subsection{Solve the PDE}
Each solver has a \verb|compile| and \verb|fit| method, to give the package a feel similar to modern popular machine learning or deep learning frameworks such as Keras~\citep{chollet2015keras} or scikit-learn~\citep{pedregosa2011scikit}. In most instances, the \verb|compile| function places parameters such as the domain size and shape, neural network sizes, BCs/ICs, etc. into the solver, and the \verb|fit| function takes only the number of iterations of Keras optimizer runs or newton solver runs.

A feature unique to \textit{TensorDiffEq} is that the Keras neural network model can be exported and saved for later use. This could allow for training on a data center platform, but inference on a local machine. Additionally, being able to export the Keras neural network model opens the door to transfer learning possibilities previously difficult with the versions of Neural PDE solvers currently in circulation. In the case of \textit{TensorDiffEq}, this is a natural result of leaning on the Tensorflow/Keras APIs. 

\section{Solving Inverse Problems}
\textit{TensorDiffEq} comes with a base class solver for inverse problems. Inverse problems can imply parameter estimation or even estimate the interactions between nonlinear operators~\citep{lu2019deeponet} from data. 
\textit{TensorDiffEq} contains solvers that perform parameter estimation in a PDE system. These parameters can be mobility parameters, diffusivity parameters, etc, where there is some level of a priori physical knowledge about the system in question, but a specific parameter may be unknown. \textit{TensorDiffEq} contains built-in support for solving of such systems that can be solved in $N$D cases. Parameters are defined as variables that are learned over the course of the training, therefore a natural output is a trained $u(\textbf{X}, t)$ solution as well as the estimate of the parameters in question.

\section{Conclusion}
In this article, the authors introduce \textit{TensorDiffEq}, a scalable multi-GPU solver for PINNs/NeuralPDEs. Some of the main highlights of the software are covered, and more features are currently underway. \textit{TensorDiffEq} contains support for various types of initial conditions, boundary conditions, and allows the user to custom-define their PDE system for their specific problem. In the event that inverse modeling is required, \textit{TensorDiffEq} contains solvers that will accommodate parameter estimation of a PDE system. Currently, \textit{TensorDiffEq} is the only software suite to support self-adaptive solving, demonstrated to improve training convergence and accuracy of the final solution. \textit{TensorDiffEq} takes a step forward in modern implementations of PINN solvers, and fills a unique niche of being a fully open-source multi-GPU PINN solver in the current ecosystem of Scientific Machine Learning software offerings.

% Acknowledgements should go at the end, before appendices and references

\acks{The authors would like to acknowledge the support of the D$^3$EM program funded through NSF Award DGE-1545403. The authors would further like to thank the US Army CCDC Army Research Lab for their generous support and affiliation, as well as the Nvidia DGX Station hardware support, which allowed the development of the software highlighted in this publication. }


\newpage

\bibliography{sample}

\begin{thebibliography}{24}
\providecommand{\natexlab}[1]{#1}
\providecommand{\url}[1]{\texttt{#1}}
\expandafter\ifx\csname urlstyle\endcsname\relax
  \providecommand{\doi}[1]{doi: #1}\else
  \providecommand{\doi}{doi: \begingroup \urlstyle{rm}\Url}\fi

\bibitem[Abadi et~al.(2015)Abadi, Agarwal, Barham, Brevdo, Chen, Citro,
  Corrado, Davis, Dean, Devin, Ghemawat, Goodfellow, Harp, Irving, Isard, Jia,
  Jozefowicz, Kaiser, Kudlur, Levenberg, Man\'{e}, Monga, Moore, Murray, Olah,
  Schuster, Shlens, Steiner, Sutskever, Talwar, Tucker, Vanhoucke, Vasudevan,
  Vi\'{e}gas, Vinyals, Warden, Wattenberg, Wicke, Yu, and
  Zheng]{tensorflow2015-whitepaper}
Mart\'{\i}n Abadi, Ashish Agarwal, Paul Barham, Eugene Brevdo, Zhifeng Chen,
  Craig Citro, Greg~S. Corrado, Andy Davis, Jeffrey Dean, Matthieu Devin,
  Sanjay Ghemawat, Ian Goodfellow, Andrew Harp, Geoffrey Irving, Michael Isard,
  Yangqing Jia, Rafal Jozefowicz, Lukasz Kaiser, Manjunath Kudlur, Josh
  Levenberg, Dan Man\'{e}, Rajat Monga, Sherry Moore, Derek Murray, Chris Olah,
  Mike Schuster, Jonathon Shlens, Benoit Steiner, Ilya Sutskever, Kunal Talwar,
  Paul Tucker, Vincent Vanhoucke, Vijay Vasudevan, Fernanda Vi\'{e}gas, Oriol
  Vinyals, Pete Warden, Martin Wattenberg, Martin Wicke, Yuan Yu, and Xiaoqiang
  Zheng.
\newblock {TensorFlow}: Large-scale machine learning on heterogeneous systems,
  2015.
\newblock URL \url{http://tensorflow.org/}.
\newblock Software available from tensorflow.org.

\bibitem[Abadi et~al.(2016)Abadi, Barham, Chen, Chen, Davis, Dean, Devin,
  Ghemawat, Irving, Isard, et~al.]{abadi2016tensorflow}
Mart{\'\i}n Abadi, Paul Barham, Jianmin Chen, Zhifeng Chen, Andy Davis, Jeffrey
  Dean, Matthieu Devin, Sanjay Ghemawat, Geoffrey Irving, Michael Isard, et~al.
\newblock Tensorflow: A system for large-scale machine learning.
\newblock In \emph{12th $\{$USENIX$\}$ symposium on operating systems design
  and implementation ($\{$OSDI$\}$ 16)}, pages 265--283, 2016.

\bibitem[Allen and Cahn(1972)]{allen1972ground}
Samuel~Miller Allen and John~W Cahn.
\newblock Ground state structures in ordered binary alloys with second neighbor
  interactions.
\newblock \emph{Acta Metallurgica}, 20\penalty0 (3):\penalty0 423--433, 1972.

\bibitem[Baker et~al.(2019)Baker, Alexander, Bremer, Hagberg, Kevrekidis, Najm,
  Parashar, Patra, Sethian, Wild, Willcox, and Lee]{osti_1478744}
Nathan Baker, Frank Alexander, Timo Bremer, Aric Hagberg, Yannis Kevrekidis,
  Habib Najm, Manish Parashar, Abani Patra, James Sethian, Stefan Wild, Karen
  Willcox, and Steven Lee.
\newblock Workshop report on basic research needs for scientific machine
  learning: Core technologies for artificial intelligence, 2 2019.

\bibitem[Baydin et~al.(2017)Baydin, Pearlmutter, Radul, and
  Siskind]{baydin2017automatic}
At{\i}l{\i}m~G{\"u}nes Baydin, Barak~A Pearlmutter, Alexey~Andreyevich Radul,
  and Jeffrey~Mark Siskind.
\newblock Automatic differentiation in machine learning: a survey.
\newblock \emph{The Journal of Machine Learning Research}, 18\penalty0
  (1):\penalty0 5595--5637, 2017.

\bibitem[Chollet et~al.(2015)]{chollet2015keras}
Fran\c{c}ois Chollet et~al.
\newblock Keras.
\newblock \url{https://keras.io}, 2015.

\bibitem[Dissanayake and Phan-Thien(1994)]{dissanayake1994neural}
MWMG Dissanayake and N~Phan-Thien.
\newblock Neural-network-based approximations for solving partial differential
  equations.
\newblock \emph{communications in Numerical Methods in Engineering},
  10\penalty0 (3):\penalty0 195--201, 1994.

\bibitem[Haghighat and Juanes(2021)]{haghighat2021sciann}
Ehsan Haghighat and Ruben Juanes.
\newblock Sciann: A keras/tensorflow wrapper for scientific computations and
  physics-informed deep learning using artificial neural networks.
\newblock \emph{Computer Methods in Applied Mechanics and Engineering},
  373:\penalty0 113552, 2021.

\bibitem[Hennigh et~al.(2020)Hennigh, Narasimhan, Nabian, Subramaniam,
  Tangsali, Rietmann, Ferrandis, Byeon, Fang, and Choudhry]{hennigh2020nvidia}
Oliver Hennigh, Susheela Narasimhan, Mohammad~Amin Nabian, Akshay Subramaniam,
  Kaustubh Tangsali, Max Rietmann, Jose del~Aguila Ferrandis, Wonmin Byeon,
  Zhiwei Fang, and Sanjay Choudhry.
\newblock Nvidia simnet\^{}$\{$TM$\}$: an ai-accelerated multi-physics
  simulation framework.
\newblock \emph{arXiv preprint arXiv:2012.07938}, 2020.

\bibitem[Kingma and Ba(2014)]{kingma2014adam}
Diederik~P Kingma and Jimmy Ba.
\newblock Adam: A method for stochastic optimization.
\newblock \emph{arXiv preprint arXiv:1412.6980}, 2014.

\bibitem[Lagaris et~al.(1998)Lagaris, Likas, and
  Fotiadis]{lagaris1998artificial}
Isaac~E Lagaris, Aristidis Likas, and Dimitrios~I Fotiadis.
\newblock Artificial neural networks for solving ordinary and partial
  differential equations.
\newblock \emph{IEEE transactions on neural networks}, 9\penalty0 (5):\penalty0
  987--1000, 1998.

\bibitem[Lu et~al.(2019)Lu, Jin, and Karniadakis]{lu2019deeponet}
Lu~Lu, Pengzhan Jin, and George~Em Karniadakis.
\newblock Deeponet: Learning nonlinear operators for identifying differential
  equations based on the universal approximation theorem of operators.
\newblock \emph{arXiv preprint arXiv:1910.03193}, 2019.

\bibitem[Lu et~al.(2021)Lu, Meng, Mao, and Karniadakis]{lu2021deepxde}
Lu~Lu, Xuhui Meng, Zhiping Mao, and George~Em Karniadakis.
\newblock {DeepXDE}: A deep learning library for solving differential
  equations.
\newblock \emph{SIAM Review}, 63\penalty0 (1):\penalty0 208--228, 2021.
\newblock \doi{10.1137/19M1274067}.

\bibitem[McClenny and Braga-Neto(2020)]{mcclenny2020self}
Levi McClenny and Ulisses Braga-Neto.
\newblock Self-adaptive physics-informed neural networks using a soft attention
  mechanism.
\newblock \emph{arXiv preprint arXiv:2009.04544}, 2020.

\bibitem[Paszke et~al.(2017)Paszke, Gross, Chintala, Chanan, Yang, DeVito, Lin,
  Desmaison, Antiga, and Lerer]{paszke2017automatic}
Adam Paszke, Sam Gross, Soumith Chintala, Gregory Chanan, Edward Yang, Zachary
  DeVito, Zeming Lin, Alban Desmaison, Luca Antiga, and Adam Lerer.
\newblock Automatic differentiation in pytorch.
\newblock 2017.

\bibitem[Pedregosa et~al.(2011)Pedregosa, Varoquaux, Gramfort, Michel, Thirion,
  Grisel, Blondel, Prettenhofer, Weiss, Dubourg, et~al.]{pedregosa2011scikit}
Fabian Pedregosa, Ga{\"e}l Varoquaux, Alexandre Gramfort, Vincent Michel,
  Bertrand Thirion, Olivier Grisel, Mathieu Blondel, Peter Prettenhofer, Ron
  Weiss, Vincent Dubourg, et~al.
\newblock Scikit-learn: Machine learning in python.
\newblock \emph{Journal of machine learning research}, 12\penalty0
  (Oct):\penalty0 2825--2830, 2011.

\bibitem[Rackauckas and Nie(2017)]{DifferentialEquations.jl-2017}
Christopher Rackauckas and Qing Nie.
\newblock Differentialequations.jl – a performant and feature-rich ecosystem
  for solving differential equations in julia.
\newblock \emph{The Journal of Open Research Software}, 5\penalty0 (1), 2017.
\newblock \doi{10.5334/jors.151}.
\newblock URL \url{https://app.dimensions.ai/details/publication/pub.1085583166
  and
  http://openresearchsoftware.metajnl.com/articles/10.5334/jors.151/galley/245/download/}.
\newblock Exported from https://app.dimensions.ai on 2019/05/05.

\bibitem[Raissi(2018)]{raissi2018forward}
Maziar Raissi.
\newblock Forward-backward stochastic neural networks: Deep learning of
  high-dimensional partial differential equations.
\newblock \emph{arXiv preprint arXiv:1804.07010}, 2018.

\bibitem[Raissi et~al.(2019)Raissi, Perdikaris, and
  Karniadakis]{raissi2019physics}
Maziar Raissi, Paris Perdikaris, and George~E Karniadakis.
\newblock Physics-informed neural networks: A deep learning framework for
  solving forward and inverse problems involving nonlinear partial differential
  equations.
\newblock \emph{Journal of Computational Physics}, 378:\penalty0 686--707,
  2019.

\bibitem[Revels et~al.(2016)Revels, Lubin, and Papamarkou]{revels2016forward}
Jarrett Revels, Miles Lubin, and Theodore Papamarkou.
\newblock Forward-mode automatic differentiation in julia.
\newblock \emph{arXiv preprint arXiv:1607.07892}, 2016.

\bibitem[Wang et~al.(2020{\natexlab{a}})Wang, Teng, and
  Perdikaris]{wang2020understanding}
Sifan Wang, Yujun Teng, and Paris Perdikaris.
\newblock Understanding and mitigating gradient pathologies in physics-informed
  neural networks.
\newblock \emph{arXiv preprint arXiv:2001.04536}, 2020{\natexlab{a}}.

\bibitem[Wang et~al.(2020{\natexlab{b}})Wang, Yu, and Perdikaris]{wang2020and}
Sifan Wang, Xinling Yu, and Paris Perdikaris.
\newblock When and why pinns fail to train: A neural tangent kernel
  perspective.
\newblock \emph{arXiv preprint arXiv:2007.14527}, 2020{\natexlab{b}}.

\bibitem[Wight and Zhao(2020)]{wight2020solving}
Colby~L Wight and Jia Zhao.
\newblock Solving allen-cahn and cahn-hilliard equations using the adaptive
  physics informed neural networks.
\newblock \emph{arXiv preprint arXiv:2007.04542}, 2020.

\bibitem[Zeiler(2012)]{zeiler2012adadelta}
Matthew~D Zeiler.
\newblock Adadelta: an adaptive learning rate method.
\newblock \emph{arXiv preprint arXiv:1212.5701}, 2012.

\end{thebibliography}

\end{document}